\documentclass[nofootinbib,prd,preprintnumbers,showpacs,superscriptaddress,twocolumn]{revtex4}

\usepackage{amssymb}
\usepackage{amsmath}
\usepackage{amsthm}
\usepackage[english]{babel}
\usepackage{exscale}
\usepackage[T1]{fontenc}
\usepackage[final]{graphicx}
\usepackage{subfigure}
\usepackage[latin9]{inputenc}
\usepackage{mathpazo}
\usepackage{psfrag}
\usepackage{textcomp}

\begin{document}

\title{Non-symmetric trapped surfaces in the Schwarzschild and Vaidya
  spacetimes} 

\author{Erik Schnetter}
\email{schnetter@cct.lsu.edu}
\homepage{http://www.cct.lsu.edu/about/focus/numerical/}
\homepage{http://numrel.aei.mpg.de/}

\affiliation{Center for Computation and Technology,
302 Johnston Hall, Louisiana State University, Baton Rouge, LA 70803, USA}

\affiliation{Max-Planck-Institut für Gravitationsphysik,
Albert-Einstein-Institut, Am Mühlenberg 1, D-14476 Golm, Germany}

\author{Badri Krishnan}
\email{badri.krishnan@aei.mpg.de}
\homepage{http://www.aei.mpg.de/}

\affiliation{Max-Planck-Institut für Gravitationsphysik,
Albert-Einstein-Institut, Am Mühlenberg 1, D-14476 Golm, Germany}

\date{January 2, 2006}

\begin{abstract}

  Marginally trapped surfaces (MTSs) are commonly used in numerical
  relativity to locate black holes.  For dynamical black holes, it is
  not known generally if this procedure is sufficiently reliable.
  Even for Schwarzschild black holes, Wald and Iyer constructed
  foliations which come arbitrarily close to the singularity but do
  not contain any MTSs.  In this paper, we review the Wald--Iyer
  construction, discuss some implications for numerical relativity,
  and generalize to the well known Vaidya spacetime describing
  spherically symmetric collapse of null dust. In the Vaidya
  spacetime, we numerically locate non-spherically symmetric trapped
  surfaces which extend outside the standard spherically symmetric
  trapping horizon.  This shows that MTSs are common in this spacetime
  and that the event horizon is the most likely candidate for the
  boundary of the trapped region.

\end{abstract}

\preprint{AEI-2005-161}
\maketitle

\noindent \textbf{Introduction:} In stationary black hole spacetimes,
there is a strong correspondence between marginally trapped surfaces
(MTSs) and event horizons (EHs) because cross sections of stationary
EHs are MTSs.  MTSs also feature prominently in the frameworks of
isolated, dynamical and trapping horizons which have shed considerable
light on the properties of classical and quantum black holes even in
the dynamical regime.  See e.g.\
\cite{Ashtekar:2004cn,Gourgoulhon-Jaramillo-Review,Booth-review} for
reviews.  Numerical simulations routinely look for MTSs to locate
black holes on a Cauchy surface.  This is because, while MTSs can be
located on a Cauchy surface in real time, the EH can only be located
\emph{a posteriori} after the simulation has been successfully
completed.  MTSs can be useful for extracting physical information
about a black hole in a numerical simulation \cite{Dreyer02a}.

However, in dynamical situations, the correspondence between MTSs and
EHs is lost (beyond the fact that MTSs are enclosed by the EH); the
event horizon is, in general, an expanding null surface, while
outgoing light rays from a MTS have, by definition, zero expansion.
An explicit example was constructed by Wald and Iyer \cite{Wald92}
where they showed that even in the Schwarzschild spacetime, there
exist perfectly regular Cauchy surfaces which come arbitrarily close
to the singularity and foliate the spacetime, but which nevertheless
do not contain any MTSs.  While this has not been an issue in most
numerical simulations to date, it raises the question of whether MTSs
can be found generally in numerical simulations of black hole
spacetimes.

However, there are other results which indicate that MTSs should be
common in black hole spacetimes.  For example, it was suggested by
Eardley \cite{Eardley98}, that an MTS can be locally perturbed in an
arbitrary spacelike direction to yield a 1-parameter family of MTSs.
A precise formulation of this idea and its proof follows from recent
results of Andersson et al.\ \cite{ams05}.  From this result, it seems
plausible that for a ``generic'' Cauchy surface $\Sigma$ passing
through the black hole region, one should be able to perturb a nearby
MTS to lie on $\Sigma$.  Thus, Cauchy surfaces passing through the
black hole should ``generically'' contain a MTS.  A related issue is
the question of the boundary of the trapped region of a black hole
spacetime; clearly, a MTS cannot be perturbed to lie outside this
boundary.  Eardley suggests that the boundary should be the event
horizon, while arguments by Hayward suggest that the boundary should
be a trapping horizon \cite{Hayward94a}.  For stationary black holes
the two notions coincide, but not for dynamical black holes.

In this paper, we study these issues from a numerical relativity
perspective.  In particular, we study the spherically symmetric
Schwarzschild and Vaidya spacetimes using foliations not adapted to
the spherical symmetry.  Very little is known so far about trapped or
marginally trapped surfaces on such slices, either analytically or
numerically.  We find that for Vaidya, MTSs are indeed easy to locate,
and we do not encounter the problem suggested by Wald and Iyer.  This
suggests that Cauchy surfaces of the type suggested by Wald and Iyer
(which presumably exist also in the Vaidya spacetime) are exceptional.
In Vaidya, we find that the non-symmetric marginal surfaces lie
partially outside the standard $r=2M$ surface (which in this case is a
trapping horizon, and lies inside the event horizon), thus indicating
that the event horizon is a better candidate for the boundary of the
trapped region as suggested by Eardley.  In the remainder of this
paper, we review basic concepts regarding trapped surfaces and
horizons, outline Wald and Iyer's construction with an explicit
example and remark on its implications for numerical relativity, and
finally discuss trapped surfaces in the Vaidya spacetime.
\vspace{2.0ex}

\noindent\textbf{Trapped surfaces and horizons:} Let $\ell^a$ and
$n^a$ be the future directed null-normals of a closed 2-surface $S$.
Let $q_{ab}$ be the 2-metric on $S$ induced by the spacetime metric.
The expansion of $\ell^a$ is $\Theta_{(\ell)} = q^{ab}\nabla_a\ell_b$
with a similar definition for $\Theta_{(n)}$.  The surface $S$ is said
to be trapped if both expansions are negative: $\Theta_{(\ell)} < 0$
and $\Theta_{(n)} < 0$.  For a \emph{marginally trapped surface}
(MTS), one or both of these inequalities are replaced by an equality
instead.  Weakly trapped surfaces have $\Theta_{(n)} \leq 0$,
$\Theta_{(\ell)} \leq 0$.  All these definitions are invariant under
arbitrary positive rescalings of $\ell^a$ and $n^a$.

The definition of a \emph{marginally outer-trapped surface} (MOTS)
requires a choice of an ``outgoing'' direction with respect to future
null infinity or spatial infinity.  This choice of outgoing direction
breaks the symmetry between the two null normals $\ell^a$ and $n^a$.
A MOTS is thus a MTS with $\Theta_{(\ell)}=0$, where $\ell$ is the
outgoing direction.

The \emph{trapped region} is the region where trapped surfaces exist.
This is defined either in the full spacetime or on a Cauchy surface
$\Sigma$. A point is in the trapped region if there is a trapped
surface which contains that point.  Similarly, a point is in the
trapped region of $\Sigma$ iff there is a trapped surface on $\Sigma$
that contains this point.  The \emph{apparent horizon} (AH) on
$\Sigma$ is the boundary of the trapped region of $\Sigma$.  As such,
its definition is so complicated that it is numerically not
feasible to look for it directly.  However, an AH is also a MTS
\cite{Kriele97}, and these can be efficiently detected.

Finally, a \emph{marginal surface} (MS) \cite{Hayward94a} is a surface
where one of the null normal's expansion vanishes, i.e.,
$\Theta_{(\ell)} = 0$ where $\ell$ can be any of the two null
directions, \emph{with no restriction on} $\Theta_{(n)}$.  Unlike the
definition of a MOTS, marginal surfaces do not require globally
defined outgoing/ingoing directions.  It is called a \emph{future
  marginal surface} if $\Theta_{(n)}<0$ and a \emph{past marginal
  surface} if $\Theta_{(n)}>0$.  A future marginal surface is the same
as a MTS and usually arises in numerical simulations as the
cross-section of a \emph{dynamical horizon} (DH) \cite{Ashtekar03a} or
an \emph{isolated horizon} (IH) \cite{Dreyer02a}, or more generally, a
trapping horizon \cite{Hayward94a}.
\vspace{2.0ex}

\noindent\textbf{The Wald--Iyer construction:}
Wald and Iyer \cite{Wald92} construct a foliation of the extended
Schwarzschild spacetime in which the spacelike hypersurfaces come
arbitrarily close to the singularity, but nevertheless do not contain
any trapped surfaces.  It should be noted that these foliations, while
special, are not pathological in any sense and they can be readily
constructed in a numerical code.  Wald and Iyer prove that, if the
intersection of the slice with the trapped region lies in the past of,
roughly speaking, ``a single event on the future singularity'', no
slice of such a foliation contains a trapped surface.  This
construction relies on the existence of angular horizons in the black
hole region, just as in a cosmological spacetime near the initial
singularity.

\begin{figure}
  \psfrag{I+}{$\mathcal{I}^+$}
  \psfrag{i+}{$i^+$}
  \psfrag{NP}{NP} 
  \psfrag{SP}{SP}
  \psfrag{i0}{$i^0$}
  \psfrag{ell}{$\ell^a$}
  \psfrag{n}{$n^a$}
  \includegraphics[width=0.35\textwidth]{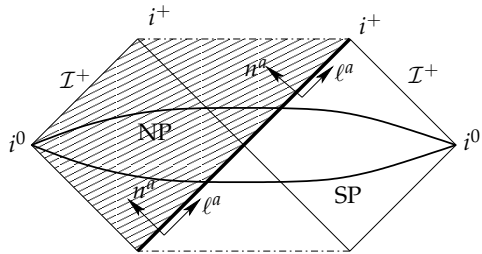}
  \caption{Penrose diagram of the extended Schwarzschild spacetime,
    from a numerical relativity point of view.  The bold line is the
    null world tube of marginal surfaces.  The hatched area is the
    region of spacetime that is invisible to the ``observer at
    infinity'', which is located at the $\mathcal{I}^+$ to the right.
    Every point in this figure is a sphere as in conventional Penrose
    diagrams, \emph{except} for the curves labeled NP and SP. If we
    represent the Cauchy surface as $f(T,X,\theta,\phi) = 0$, then NP
    and SP are respectively the projections of the north pole
    ($\theta=0$) and south pole ($\theta=\pi$); the intermediate
    angles lie in between.  NP enters the trapped region, but SP does
    not. }
  \label{fig:penrose-numrel}
\end{figure}

An explicit example of one such Cauchy surface is easy to construct.
Consider the extended Schwarzschild spacetime in Kruskal coordinates
$(T,X,\theta,\phi)$ (see eq.\ (6.4.29) of \cite{Wald84}).  The
hypersurface $T = k\, \cos\theta$ can be easily shown to satisfy the
Wald--Iyer condition for $k<1/2$.  Thus, even though this surface
enters the black hole region, it does not contain any trapped
surfaces.  Such a slice is depicted schematically in figure
\ref{fig:penrose-numrel}.  It intersects the black hole horizon for
$T>0$, and the white hole horizon for $T<0$.

Even though it is a fact that the above Cauchy surface does not
contain a MOTS, standard apparent horizon trackers employed in
numerical simulations will happily find an ``apparent horizon'' on
this slice.  This apparent contradiction is an issue of terminology.
What the apparent horizon tracker will locate is the intersection of
the Cauchy surface with the surface $T=X$, which is the bold line in
figure \ref{fig:penrose-numrel}.  The intersection is the 2-sphere
given by $T = X = k\, \cos\theta$.  In numerical relativity, one
typically chooses that part of $\mathcal{I}^+$ which belongs to one
specific asymptotically flat end of the spacetime.  Thus, the
``outgoing'' null normal $\ell^a$ and the ingoing null normal $n^a$
are the ones shown in figure \ref{fig:penrose-numrel}.  With this
choice of $\ell^a$, the surface given above satisfies $\Theta_{(\ell)}
= 0$.  However, this ``apparent horizon'' is not a MTS because
$\Theta_{(n)} < 0$ on the black hole portion and $\Theta_{(n)} >0$ on
the white hole portion.

What is often loosely called ``apparent horizon'' in numerical
relativity, or almost as loosely ``marginally outer-trapped surface'',
is really only a marginal surface, or a future marginal surface if the
condition $\Theta_{(n)} < 0$ is checked (which it is often not).
Determining the globally outgoing direction is usually either
unpractical or impossible in numerical simulations.  If done, it
requires some additional knowledge of the simulated spacetime that the
code itself generally does not have.  The easiest way to avoid such
situations in numerical relativity is to explicitly make sure that the
apparent horizon is future trapped by verifying that $\Theta_{(n)}$ is
negative everywhere, both in the initial data and during evolution.
Regarding black hole initial data, the construction presented in
\cite{Dain04} will ensure that the marginal surfaces are future
trapped and will therefore avoid any of the Wald--Iyer slices
if the lapse function is kept nonnegative everywhere.
\vspace{2.0ex}

\noindent\textbf{Trapped surfaces in the Vaidya spacetime:}
Generalizing to dynamical situations, consider the Vaidya spacetime
which describes spherically symmetric collapse of null dust
(radiation) \cite{Vaidya51a}.  This is an astrophysically unrealistic
toy model, but it does serve as a very useful testing ground. It has
been extensively used, for example, to study the formation of naked
singularities.  In ingoing Eddington--Finkelstein coordinates
$(v,r,\theta,\phi)$, the metric is
\begin{eqnarray}
  \label{eq:vaidya}
   ds^2 & = & - \left( 1 - \frac{2M(v)}{r} \right) dv^2 + 2\, dv\, dr
   + r^2\, d\Omega^2\,,
\end{eqnarray}
where the mass function $M(v)$ can be specified as a function of the
null coordinate $v$. For constant $M(v)$, this is just the
standard Schwarzschild metric in ingoing Eddington--Finkelstein
coordinates. The stress energy tensor is determined by the derivative
of $M(v)$:
\begin{eqnarray}
  T_{ab} & = & \frac{\dot M(v)}{4 \pi r^2} \partial_a v\, \partial_b v
\end{eqnarray}
where $\dot M = \partial M / \partial v \geq 0$.  We shall use a time
coordinate defined as $t=v-r$ and we shall take the mass
function to be non-zero only for $v>0$.  Thus, the spacetime is flat
for $v\leq 0$. 

It is easy to see that the only spherically symmetric MTSs
are the spheres given
by $r=2M(v_0)$ for a specified $v_0$.  These will be the apparent
horizons on spherically symmetric Cauchy surfaces which intersect the
$r=2M(v)$ surface.  Let us denote the $r=2M(v)$ surface by $H$. It is
easy to show that $H$ is spacelike and is a trapping horizon.  The EH
lies outside $H$ and is strictly separated from $H$ when $\dot M > 0$;
at late times, $H$ asymptotes to the EH \cite{Ashtekar03a}.

Let us now consider non-spherically symmetric Cauchy surfaces.  There
is now an important qualitative difference from the Schwarzschild
case. There, the analog of $H$ was null and expansion free; the
intersection of any spacelike surface with $H$ was then a marginal
surface, as long as this intersection was, topologically, a complete
sphere.  This is also true more generally when the black hole is
isolated in an otherwise dynamical spacetime (if $H$ is an isolated
horizon).  However, in genuinely dynamical situations, $H$ is
spacelike as in the Vaidya example.  In this case, if the intersection
of a Cauchy surface with $H$ is not one of the spherically symmetric
marginal surfaces, \emph{then the intersection cannot be a marginal
  surface even if it is a complete 2-sphere}.  This statement follows
directly from Theorem 4.2 of \cite{Ashtekar05}.  Thus the question
naturally arises: are there apparent horizons on non-symmetric Cauchy
surfaces in the Vaidya spacetime?  One would expect there to be
Wald--Iyer Cauchy surfaces which come arbitrarily close to the
singularity but which do not contain any marginal surfaces, but we
shall now see that apparent horizons do exist on a large class of
non-symmetric Cauchy surfaces.

We choose the mass function 
\begin{eqnarray}
   M(v) & = & \left\{
\begin{array}{ll}
0 & \textrm{for}\, v \le 0
\\
M_0\, v^2 / (v^2 + W^2) & \textrm{for}\, v > 0
\end{array}
   \right.
\end{eqnarray}
with the constants $M_0=1$ and $W=1/10$.  This is a short pulse of
radiation that forms a black hole with the final mass $M_0$. This mass
function is only $C^1$ at $v=0$, but our results are unchanged
qualitatively for other mass functions with higher differentiability.
For this mass function the singular point $v=0$, $r=0$ is locally
naked (see e.g.\ \cite{Kuroda84a}), but this is not relevant for our
purposes. 

We examine the spacetime with a slicing that is only axially
symmetric.  We use a time coordinate $\bar t$ given by
\begin{eqnarray}
  \label{eq:tbar}
  \bar{t} & = & t - \alpha z = v - r\,(1 +\alpha \cos\theta)\,,
\end{eqnarray}
where $t = v-r$ is the standard Vaidya time, and the constant
$\alpha$ determines how much the slice is boosted in the $z$
direction.  We chose $\alpha = 10/11$. We have also examined other more
complicated foliations, but the results presented below do not change
qualitatively.

\begin{figure}
  \includegraphics[width=0.4\textwidth]{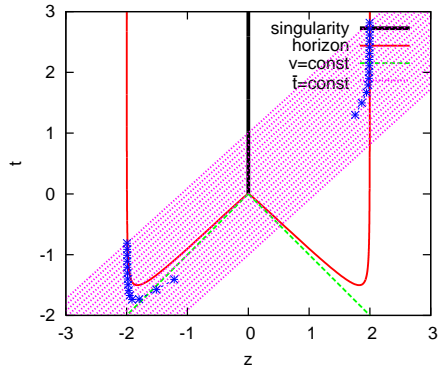}
  \caption{The $t$-$z$ section of the Vaidya spacetime where $t=v-r$.
    The tilted lines are (sections of) the axisymmetric surfaces given
    in eq.\ (\ref{eq:tbar}) for a range of $\bar t$ values.  The
    marginal surfaces on these sections are marked by a ``$\star$''.
    The dynamical horizon $H$ (the $r=2M(v)$ surface) are the pair of
    bold curves, and the two dashed straight lines are the $v=0$ light
    cone which is the boundary of the flat portion of the spacetime.
    The singularity is the positive $t$-axis ($z=0$, $t\geq 0$). }
  \label{fig:tilted-slices}
\end{figure}
\begin{figure}[htbp]
  \begin{center}
    \mbox{
      \subfigure[$\bar t=-0.3$]{\scalebox{0.7}{\label{fig:t-0.3}\includegraphics{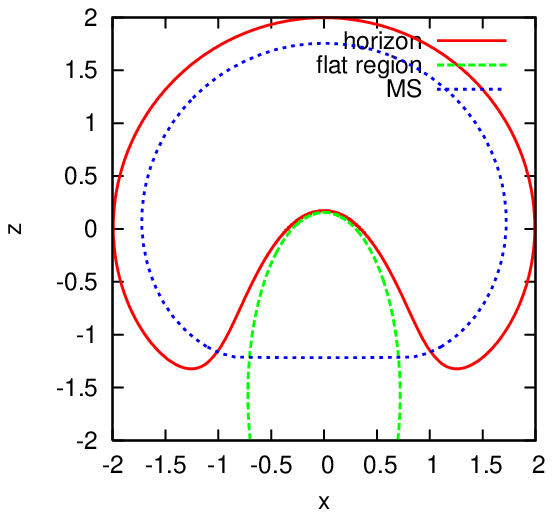}}} \hspace{-0.5in}
      \subfigure[$\bar t=0.0$]{\scalebox{0.7}{\label{fif:t0.0}\includegraphics{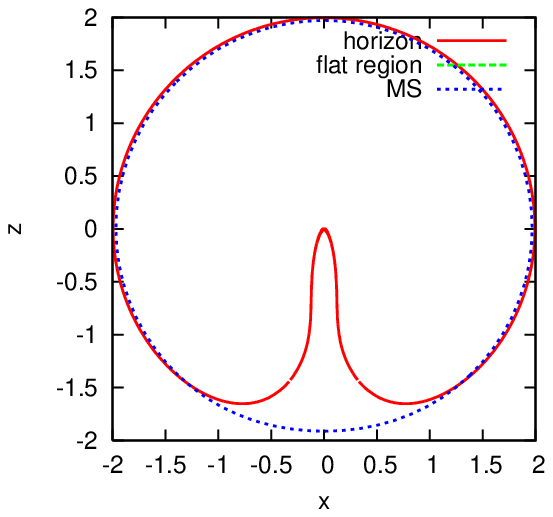}}} 
      }
      \caption{ The location of the distorted marginal surface on the
        $x$-$z$ section of the Cauchy surfaces at $\bar t=-0.3,0.0$.
        The solid curve is the intersection with the dynamical horizon
        $H$, the dotted curve is the distorted MS, and the region
        inside the dashed curve is the $v<0$ region, i.e., the flat
        portion of spacetime.  Note that at $\bar t=0.0$, the Cauchy
        surface does not intersect the flat portion at all, and the
        distorted MS coincides closely with the DH except near its
        south pole.  In both cases, the north pole is inside the DH
        while the south pole is outside. }
    \label{fig:distorted-ms}
  \end{center}
\end{figure}
The results are shown in figures \ref{fig:tilted-slices} and
\ref{fig:distorted-ms}.  Figure \ref{fig:tilted-slices} shows the
$t$-$z$ section of the Cauchy surfaces for a range of $\bar t$ values,
and the distorted MSs on these sections.  This clearly shows that the
MSs extend outside $H$ and can also extend into the flat region.
Figure \ref{fig:distorted-ms} shows the MS on the $\bar t=-0.3$ and
$\bar t=0$ slices.  The MS at $\bar t=0$ is a future marginally
trapped surface, i.e., $\Theta_{(n)}< 0$.  At $\bar t=-0.3$, the MS
extends into the flat portion, and this part of $S$ is planar, with
$\Theta_{(n)} = \Theta_{(\ell)} = 0$. On the rest of the sphere,
$\Theta_{(n)} < 0$ as expected; this is therefore a \emph{weakly}
marginally trapped surface.  Furthermore, the 3-dimensional world tube
obtained by stacking up all the MSs turns out to be spacelike; the
ones with $\Theta_{(n)} < 0$ form a dynamical horizon.

We also look for surfaces with a small non-vanishing expansion
$\Theta_{(\ell)} = \pm 10^{-3}$.  These surfaces can be viewed as
radial deformations of the MS; the outward deformation has
$\Theta_{(\ell)} > 0$, and $\Theta_{(\ell)}<0$ for the inward
deformation.  At $\bar t=0$, the inward deformation is strictly
trapped and the outward deformation is strictly untrapped.  At $\bar
t=-0.3$, the inward deformation has $\Theta_{(n)} > 0$ in the flat
region and $\Theta_{(n)} <0$ elsewhere.  The outward deformation has
$\Theta_{(n)} < 0$ everywhere and is thus strictly untrapped.  We have
not been able to find strictly trapped surfaces which extend into the
flat portion of spacetime.  Sufficiently far in the future, the MSs
asymptote to the spherically symmetric DH and also come arbitrarily
close to the EH.  Finally, there are restrictions on the location of
trapped surfaces in the presence of a dynamical horizon
\cite{Ashtekar05}. We have verified that these restrictions are
satisfied.  The existence of such distorted MSs was already suggested
in \cite{Ashtekar05}, but with no restrictions on $\Theta_{(n)}$; here
we have also shown $\Theta_{(n)}\leq 0$.
\vspace{2.0ex}

\noindent\textbf{Conclusions:} We have numerically studied
non-symmetric trapped surfaces in simple spherically symmetric
spacetimes.  We have seen that the Wald--Iyer example illustrates the
importance of verifying $\Theta_{(n)} \leq 0$ for apparent horizons
located numerically.  In Vaidya, we have found trapped surfaces which
extend outside the usual $r=2M$ surface $H$.  This shows that $H$ is
not the boundary of the trapped region.  We have also found marginal
surfaces that extend into the flat region of the spacetime.

The boundary of the trapped region should be spherically symmetric,
since it is an invariantly defined geometric quantity in a spherically
symmetric spacetime.  This lends support to Eardley's conjecture that
the event horizon is the boundary of the trapped region, since the EH
is the only natural candidate.  However, we have not found strictly
trapped surfaces that extend into the flat region of the spacetime, so
that the boundary may be inside the EH.

We conclude with some open questions that need to be addressed: (i) Is
it possible to push the marginal surfaces arbitrarily close to the
event horizon, even in the flat region? This would verify that the EH
is truly the boundary of the trapped region.  (ii) For asymptotically
flat spacetimes, the event horizon is the natural candidate for the
boundary of the trapped region.  What is this boundary for
non-asymptotically flat spacetimes, e.g.\ asymptotically de~Sitter
spacetimes where the event horizon is not strictly defined?
\vspace{2.0ex}

\noindent\textbf{Acknowledgements:} We thank L.~Andersson,
A.~Ashtekar, J.~Frauendiener, G.~Galloway, S.~Hayward, M.~Korzynski,
and L.~Lehner for valuable discussions and comments.
As always, our numerical calculations would have been impossible
without the large number of people who made their work available to
the public: we used the Cactus computational toolkit
\cite{Goodale02a, cactusweb1} with a number of locally developed
thorns, J. Thornburg's apparent horizon finder AHFinderDirect
\cite{Thornburg2003:AH-finding}, and the GNU Scientific Library GSL
\cite{gslweb}.
E. Schnetter was funded by the DFG's special research centre TR-7
``Gravitational Wave Astronomy''.
This work was supported by the Albert--Einstein--Institut and the
Center for Computation \& Technology at Louisiana State University.

\bibliographystyle{apsrev-nourl}
\bibliography{bibtex/references}

\end{document}